\documentclass[11pt]{letter}
\usepackage{amsmath}
\usepackage{amssymb}
\usepackage[dvips]{graphicx}
\DeclareFontFamily{U}{msb}{}
\DeclareFontShape{U}{msb}{m}{n}{ <5> <6> <7> <8> <9> gen * msbm
        <10> <10.95> <12> <14.4> <17.28> <20.74> <24.88> msbm10}{}
\DeclareSymbolFont{AMSb}{U}{msb}{m}{n}
\DeclareMathSymbol{\realset}{\mathalpha}{AMSb}{"52}

\newcommand{\osd}[1]{\overset{\scriptscriptstyle #1}{)}}
\newcommand{\ose}[1]{\overset{\scriptscriptstyle #1}{(}}

\begin{document}

\begin{center}
\Large 
{\bf Heuristic decomposition of cones in molecular conformational space.}
\end{center}

\vspace*{6mm}
\begin{center}
{\Large Jacques Gabarro-Arpa}
\end{center}

\vspace*{6mm}
\hspace*{26mm}
Ecole Normale Sup\'erieure de Cachan, LBPA,CNRS UMR 8113      \newline
\hspace*{26mm}
61, Avenue du Pr\'esident Wilson, 94235 Cachan cedex, France  \newline

\noindent \hspace*{60mm} Email: jga@lbpa.ens-cachan.fr

\vspace*{4mm}
\hspace*{4mm} {\it Abstract}
In a previous work [physics/061108v2], it was shown that the volume spanned by a molecular system in its conformational space can be effectively bounded by a polyhedral cone, this cone is described by means of a simple combinatorial formula. On the other hand it was constructed a transversal graph structure encoding the region of conformational space accessible to the system. From the information in this graph, it is possible to decompose the main cone into a hierarchy of smaller ones that are more manageable, and are progressively more tightly bound to the region in which the system evolves.

\vspace*{3mm}
\hspace*{4mm} {\it Keywords}
{\bf Molecular Conformational Space, Polyhedral Cone, Poset, Molecular Dynamics} 



\newpage
\begin{center}
{\scshape I. Partition sequences}
\end{center}

\par
A partition of molecular conformational space (thereafter refered as $CS$) was defined and studied in [1-4] based on the notion of \textbf{dominance partition sequence}. A simple example will shows the basic ideas behind this concept. Let us have a simple four atom molecular system where an arbitrary order relation (numbering) has been defined on the set of atoms, and let their $\{x,y,z\}$ coordinates be

\begin{enumerate}
 \item $\{-17.14, \hspace{3.2mm} 43.03, \hspace{3.2mm} 0.50\}$
 \item $\{-30.35, \hspace{3.2mm} 44.92,               -2.98\}$
 \item $\{-28.22, \hspace{3.2mm} 42.82, \hspace{3.2mm} 8.89\}$
 \item $\{-21.02, \hspace{3.2mm} 37.51,               -5.19\}$ \hspace{41.7mm} (1)
\end{enumerate}

\vskip 1mm
\noindent
from (1) we see that the following relations hold :
\par
\noindent
$      x_2 \ < \ x_3 \ < \ x_4 \ < \ x_1  \ \ \ ,
 \ \ \ y_4 \ < \ y_3 \ < \ y_1 \ < \ y_2  \ \ \ ,
 \ \ \ z_4 \ < \ z_2 \ < \ z_1 \ < \ z_3$ \hspace{4mm} (2)

\par
\noindent
from relations (2) we say that, for example, coordinate $x_3$ \textbf{dominates} $x_2$ and is \textbf{dominated} by $x_4$ and $x_1$. Thus the molecular conformation defined by coordinate set (1) can be characterized by the following \textbf{dominance partition sequence} $(DPS)$

\par
\noindent
$\{\{(2)(3)(4)(1)\}_x \ ,\ \{(4)(3)(1)(2)\}_y \ , \ \{(4)(2)(1)(3)\}_z\}$ \hspace{4mm} (3)

\par
\noindent
where the atom numbers in $x$, $y$ and $z$ are ordered as in (2). From (3) we can partition the $(CS)$ of a molecular system into a set of discrete cells such that the conformations in a cell all have the same $DPS$, as was discussed in [1,4] cells in $CS$ have the shape of polyhedral cones with the vertex at the origin.

\par
The formula (3) for DPSs has the peculiar characteristic that each index is enclosed between parenthesis, this is so beause in this way it can be extended to designate sets of continguous cells. Supose we have a conformation with $DPS$ $\{... (i)(j) ...\}_c$, exchanging the coordinates $c_i$ and $c_j$ results in a new conformation with $DPS$ $\{... (j)(i) ...\}_c$ that lies in an adjacent cell, thus the  the notation for $DPS$ in (3) can be extended to

\par
\noindent
$\{... (i \ j) ...\}_c$ \hspace{73.3mm} (4)

\par
\noindent
where (4) designates all the sequences that can be obtained by a permutation of the consecutive numbers $i$ and $j$.

\par
This can be generalized [4] for any set of consecutive numbers

$\{... (i_1 i_2 \ ... \ i_n) ...\}_c$ \hspace{62.7mm} (5)

\par
where (5) designates any sequence obtained by permuting the consecutive numbers $i_1$, $i_2$ ... $i_n$. In what follows sequences like (3) will be designated as \textbf{simple} $DPS$s while (4) and (5) will be \textbf{extended} $DPS$s.

\par
An inclusion relation among $DPS$s can be defined : if $\mathcal{P}_a$ and $\mathcal{P}_b$ are two $DPS$s and if $\mathcal{S}_{\mathcal{P}_a}$ and $\mathcal{S}_{\mathcal{P}_b}$ are the sets of simple sequences that they encode, then

\par
\noindent
{\bf Definition 1}. {\it $\mathcal{P}_a   \ \subset \              \mathcal{P}_b$ \ \ if \ \
            $\mathcal{S}_{\mathcal{P}_a}  \ \subset \ \mathcal{S}_{\mathcal{P}_b}$}.

\par
with this definition extended $DPS$s amount to more than just sequences: let $\mathcal{P}$ be any $DPS$ the set $\{\mathcal{S}_{\mathcal{P}} : \forall \mathcal{P}_s \subset \mathcal{P} \Rightarrow \mathcal{P}_s \in  \mathcal{S}_{\mathcal{P}}\}$,
$\mathcal{S}_{\mathcal{P}}$ is the partially ordered set (\textbf{poset}) associated to $\mathcal{P}$ [1,6,7].

\vspace*{4mm}
\begin{center}
{\scshape II. Generalized partition sequences}
\end{center}

\par
DPSs are very useful structures: with a minimum of code they allow to designate huge numbers of cells in $CS$. However, it was shown in [5] that if dominance partition sequences are to quantify the regions enclosing molecular dynamics trajectories they require one further level of codification: parenthesis in (3) and (5) are to be allowed to overlap. For example

\par
$\{\ose{1} 3 \ 4 \
   \ose{2} 8 \ 9 \osd{1}  \
           1 \ 7 \osd{2}\}$ \hspace*{64.3mm} (6)

\par
\noindent
simultaneously encodes the extended $DPS$s $\{(3 \ 4 \ 8 \ 9)(1 \ 7)\}$ and $\{(3 \ 4)(1 \ 7 \ 8 \ 9)\}$. As it can be seen from (6) pairs of parenthesis must bear an index so we can tell the beginning and the end. In order to distinguish (6) form ordinary $DPS$s the denomination of \textbf{generalized dominance partition sequences} ($GDPS$) was proposed in [5].

\par
Coding sets of cells from conformational space as in (6) allows to define a cone in $CS$ that wraps the region where the system evolves. For example, from a 2 ns molecular dynamics trajectory of a 58 residue protein structure [9]: the pancreatic trypsin inhibitor [8], using the dominance relations matrix from the figure 1 in [5], we have calculated the minimal $GDPS$-cone enclosing the $\alpha$-carbons coordinates of the PTI

\vspace*{4mm}
\noindent
$\{\{\ose{ 1} 29 \
              48 \
              49 \
     \ose{ 2} 27 \
     \ose{ 3} 28 \
     \ose{ 4} 30 \
              31 \
              52 \
     \ose{ 5} 53 \osd{ 1} \
              47 \
              50 \osd{ 2} \
              32 \osd{ 3} \
     \ose{ 6} 26 \osd{ 4} \
              21 \
              51 \
     \ose{ 7} 19 \
              23 \
              24 \
     \ose{ 8} 20 \
              25 \
              33 \osd{ 5} \
              46 \
     \ose{ 9} 54 \
              55 \osd{ 6} \
              22 \osd{ 7} \ \newline
\hspace*{5mm} 18 \
              34 \
     \ose{10} 45 \osd{ 8} \
              17 \osd{ 9} \
     \ose{11} 44 \
     \ose{12}  5 \
     \ose{13}  8 \
     \ose{14}  6 \
              35 \osd{10} \
     \ose{15}  9 \osd{11} \
              43 \osd{12} \
              16 \
     \ose{16} 11 \osd{13} \
               7 \osd{14} \
              36 \
     \ose{17}  3 \
               4 \
              10 \osd{15} \
     \ose{18} 37 \
              42 \osd{16} \
     \ose{19} 15 \
     \ose{20} 12 \osd{17} \ \newline
\hspace*{4mm}
     \ose{21} 41 \osd{18} \
              40 \osd{19} \
              38 \
     \ose{22} 14 \osd{20} \
              13 \osd{21} \
              39 \osd{22}\}_{x}$ , \newline
\hspace*{1mm}
  $\{\ose{ 1} 15 \
              16 \
     \ose{ 2} 17 \osd{ 1} \
     \ose{ 3} 14 \osd{ 2} \
     \ose{ 4} 18 \osd{ 3} \
     \ose{ 5} 36 \
     \ose{ 6} 13 \osd{ 4} \
              37 \osd{ 5} \
     \ose{ 7} 19 \
     \ose{ 8} 34 \osd{ 6} \
              12 \
              35 \
              38 \osd{ 7} \
     \ose{ 9} 11 \osd{ 8} \
              20 \
              33 \
     \ose{10} 39 \osd{ 9} \
     \ose{11} 46 \
     \ose{12} 10 \osd{10} \
              32 \
              40 \
              47 \osd{11} \ \newline
\hspace*{4mm}
     \ose{13} 21 \osd{12} \
     \ose{14} 45 \osd{13} \
              44 \
     \ose{15} 31 \
     \ose{16}  9 \
              48 \osd{14} \
              41 \osd{15} \
     \ose{17} 22 \osd{16} \
     \ose{18} 42 \
              49 \
              50 \osd{17} \
               8 \
              30 \
              43 \
              51 \osd{18} \
     \ose{19} 23 \
     \ose{20} 24 \osd{19} \
     \ose{21}  7 \
              52 \
     \ose{22} 29 \
     \ose{23}  4 \
              53 \
              54 \osd{20} \ \newline
\hspace*{4mm}
     \ose{24} 26 \
              27 \osd{21} \
               5 \osd{22} \
               6 \
              25 \
              28 \
              55 \osd{23} \
               3 \osd{24}\}_{y}$ , \newline
\hspace*{1mm}
  $\{\ose{ 1} 26 \
     \ose{ 2} 27 \
     \ose{ 3}  8 \
              10 \
     \ose{ 4}  7 \
     \ose{ 5} 25 \
     \ose{ 6} 11 \
     \ose{ 7} 13 \osd{ 1} \
               9 \osd{ 2} \
               6 \
              24 \
     \ose{ 8} 12 \
     \ose{ 9} 28 \osd{ 3} \
              33 \osd{ 4} \
     \ose{10} 31 \osd{ 5} \
              34 \
     \ose{11} 15 \osd{ 6} \
     \ose{12} 32 \
     \ose{13} 29 \
     \ose{14} 17 \osd{ 7} \
     \ose{15} 14 \osd{ 8} \
               5 \
              23 \
               4 \ \newline
\hspace*{4mm}
     \ose{16} 22 \
              35 \
              36 \
              40 \
     \ose{17} 41 \
               3 \osd{ 9} \
     \ose{18} 30 \osd{10} \
              39 \osd{11} \
              16 \
              21 \osd{12} \
              43 \osd{13} \
              38 \osd{14} \
     \ose{19} 18 \
              19 \osd{15} \
              20 \osd{16} \
              37 \osd{17} \
              42 \
              44 \osd{18} \
     \ose{20} 55 \osd{19} \
     \ose{21} 45 \
              48 \
     \ose{22} 52 \osd{20} \
              51 \osd{21} \ \newline
\hspace*{4mm}
     \ose{23} 46 \
              47 \osd{22} \
              54 \
     \ose{24} 49 \
              53 \osd{23} \
               5 \osd{24} \}_{z}\}$ \hspace*{42.5mm} (7)

\vspace*{4mm}
\par
As stressed at the end of section I the importance of expressions like (7) lies in their associated poset: a hierarchical structure, because it allows a hierarchical decomposition of the $CS$ region into sets of smaller cones and, more important, it facilitates the sorting of the \textbf{graph of cells} [3-5] which is the structure containing all the information about cells in $CS$.

\vspace*{4mm}
\begin{center}
{\scshape III. Merging cells from adjacent nodes in the compact graph of cells}
\end{center}

\par
The \textbf{graph of cells}, or $\mathbf{G}$ is the fundamental structure of the present approach, it arose from the notion that the relative motions of small sets of atoms in the molecule can be thoroughly sampled in computer simulations. $\mathbf{G}$ encodes the set of cells in conformational space that can be accessed through the allowed combinations of these movements.

\par
This structure is constructed first by dividing the molecule into four-atom ordered sets whose $3D$-structure is that of a \textbf{simplex}\footnote{An irregular polytope with four vertices.}, next for each simplex the visited cells in its $CS$ are determined empirically from computer simulations. Then $\mathbf{G}$ is built as follows [3-4]

\begin{itemize}
 \item Each cell from a simplex is a node of $\mathbf{G}$
 \item The edges of the graph are between compatible nodes in adjacent simplexes: two cells from the $CS$s of two simplexes that share
       a face, or equivalently three atoms, are said to be \textbf{compatible} if their $DPS$s restricted to the numbers of the shared 
       atoms are equal [5].
 \item For every pair of simplexes sharing a face, each cell from the $CS$ of one simplex has an edge towards at least one cell from the
       other simplex, otherwise the structure would be geometrically inconsistent.
\end{itemize}

\textbf{Lemma 1.}
\textit{Given a simplex $\mathcal{S} \in \mathbf{G}$ and a cell $\mathbf{\xi_1} \in \mathcal{S}$ there is at least one other cell $\mathbf{\xi_2} \in \mathcal{S}$ such that their respective $DPS$s differ only by a permutation of two numbers}.

\par
Let $\mathcal{S} = \{n_{s_1},n_{s_2},n_{s_3},n_{s_4}\}$, assume that we have two cells whose coordinate $c$ $DPS$s are $\{(n_{s_1})(n_{s_2})(n_{s_3})(n_{s_4})\}_c$ and $\{(n_{s_1})(n_{s_4})(n_{s_2})(n_{s_3})\}_c$ respectively, the remaining coordinates being equal, and that there is not a cell with sequence $\{(n_{s_1})(n_{s_2})(n_{s_4})(n_{s_3})\}_c$, this is geometrically impossible because the coordinate $c_{n_{s_4}}$ cannot pass through $c_{n_{s_2}}$ without first going through $c_{n_{s_3}}$.

\par
There is a class of subgraphs of $\mathbf{G}$ called \textbf{transversals} such that they have a set of nodes consisting of one cell from every simplex and each cell has a edge towards every cell in adjacent simplexes. It was shown in [5] that each cell in a transversal is the projection of one cell from the $CS$ of the molecule, thus $\mathbf{G}$ makes possible the enumeration of accessible cells in the conformational space of a molecule.

\par
$\mathbf{G}$ can be put in a compact form called $\mathbf{C}$ by recursively agregating sets of $DPS$ in $\mathbf{G}$ into extended ones that contain them. For instance the cells from the simplex $\{ 9,10,14,15\}$ in $\mathbf{G}$ [4,5]

\par
             $\{\{\{(9)(10)(15)(14)\}_x \ , \ \{(15)(14)(10)(9)\}_y \ , \ \{(9)(10)(14)(15)\}_z\}, \newline
\hspace*{1.7mm} \{\{(9)(10)(15)(14)\}_x \ , \ \{(15)(14)(10)(9)\}_y \ , \ \{(9)(10)(15)(14)\}_z\}, \newline
\hspace*{1.7mm} \{\{(9)(10)(15)(14)\}_x \ , \ \{(15)(14)(10)(9)\}_y \ , \ \{(10)(15)(9)(14)\}_z\}, \newline
\hspace*{1.7mm} \{\{(9)(10)(15)(14)\}_x \ , \ \{(15)(14)(10)(9)\}_y \ , \ \{(10)(9)(15)(14)\}_z\}, \newline
\hspace*{1.7mm} \{\{(9)(10)(15)(14)\}_x \ , \ \{(15)(14)(10)(9)\}_y \ , \ \{(10)(9)(14)(15)\}_z\}, \newline
\hspace*{1.7mm} \{\{(9)(15)(10)(14)\}_x \ , \ \{(15)(14)(10)(9)\}_y \ , \ \{(9)(10)(14)(15)\}_z\}, \newline
\hspace*{1.7mm} \{\{(9)(15)(10)(14)\}_x \ , \ \{(15)(14)(10)(9)\}_y \ , \ \{(9)(10)(15)(14)\}_z\}, \newline
\hspace*{1.7mm} \{\{(9)(15)(10)(14)\}_x \ , \ \{(15)(14)(10)(9)\}_y \ , \ \{(10)(15)(9)(14)\}_z\}, \newline
\hspace*{1.7mm} \{\{(9)(15)(10)(14)\}_x \ , \ \{(15)(14)(10)(9)\}_y \ , \ \{(10)(9)(15)(14)\}_z\}, \newline
\hspace*{1.7mm} \{\{(9)(15)(10)(14)\}_x \ , \ \{(15)(14)(10)(9)\}_y \ , \ \{(10)(9)(14)(15)\}_z\}, \newline
\hspace*{1.7mm} \{\{(10)(9)(15)(14)\}_x \ , \ \{(15)(14)(10)(9)\}_y \ , \ \{(9)(10)(14)(15)\}_z\}, \newline
\hspace*{1.7mm} \{\{(10)(9)(15)(14)\}_x \ , \ \{(15)(14)(10)(9)\}_y \ , \ \{(9)(10)(15)(14)\}_z\}, \newline
\hspace*{1.7mm} \{\{(10)(9)(15)(14)\}_x \ , \ \{(15)(14)(10)(9)\}_y \ , \ \{(10)(15)(9)(14)\}_z\}, \newline
\hspace*{1.7mm} \{\{(10)(9)(15)(14)\}_x \ , \ \{(15)(14)(10)(9)\}_y \ , \ \{(10)(9)(15)(14)\}_z\}, \newline
\hspace*{1.7mm} \{\{(10)(9)(15)(14)\}_x \ , \ \{(15)(14)(10)(9)\}_y \ , \ \{(10)(9)(14)(15)\}_z\}\}$ \hspace{5mm} (8)

\par
can be put in a compact form in $\mathbf{C}$

\par
             $\{\{\{(9)(10 \ 15)(14)\}_x \ , \ \{(15)(14)(10)(9)\}_y \ , \ \{(10)(9 \ 15)(14)\}_z\},  \newline
\hspace*{1.7mm} \{\{(9 \ 10)(15)(14)\}_x \ , \ \{(15)(14)(10)(9)\}_y \ , \ \{(10)(9 \ 15)(14)\}_z\},  \newline
\hspace*{1.7mm} \{\{(9)(10 \ 15)(14)\}_x \ , \ \{(15)(14)(10)(9)\}_y \ , \ \{(9 \ 10)(14 \ 15)\}_z\}, \newline
\hspace*{1.7mm} \{\{(9 \ 10)(15)(14)\}_x \ , \ \{(15)(14)(10)(9)\}_y \ , \ \{(9 \ 10)(14 \ 15)\}_z\}\}$ \hspace{10.2mm} (9)

The notion of compatibility can be extended to the cells in $\mathbf{C}$: two cells in adjacent simplexes in $\mathbf{C}$ are said to be \textbf{compatible} if the dominance patterns that result from restricting their respective $DPS$ sequences to the atom numbers from the shared face are equal.

\vspace*{4mm}
\begin{center}
{\scshape IV. The fragmentation of the cone}
\end{center}

\par
The cone described in section II encloses the volume in $CS$ that the system can access, this cone has been built by looking empirically in computer simulations at the range of each atom coordinate independently [5], so part of the volume is wasted as the construction of the cone does not take into account how the $x$, $y$ and $z$ coordinates are correlated. The graph $\mathbf{C}$ can supply the missing information allowing us to create cones that wrap more closely to the cells in it.

\par
We propose here an approach to the fragmentation of the cone that is based on the following set of heuristic rules:

\begin{enumerate}
 \item For a simplex $\mathcal{S} \in \mathbf{C}$ we scan the cell sequences for each coordinate independently:
       if for the coordinate $c$ the partition sequences of two cells $ \ \mathbf{\xi_1} \ , \ \mathbf{\xi_2} \in \mathcal{S}$
       are such that $\mathcal{P}_{1,c} \ \subset \ \mathcal{P}_{2,c}$, then we set $\mathcal{P}_{1,c} \ := \ \mathcal{P}_{2,c}$.
 \item Let $\mathcal{S} = \{n_{s_1},n_{s_2},n_{s_3},n_{s_4}\}$ be a simplex in $\mathbf{C}$ and
       be $\{\mathcal{P}_{x,c} \ : 1 \leq x \leq N_{\xi}\}$ its set of cell sequences in the $c$ coordinate. 
       For 4 numbers there is a set of 24 possible simple $DPS$s, if every sequence from this set is contained in at least one
       $\mathcal{P}_{x,c}$, then $\forall x : 1 \leq x \leq N_{\xi}$ we set
       $\mathcal{P}_{x,c} \ := \ \{(n_{s_1}\ n_{s_2}\ n_{s_3}\ n_{s_4})\}_c.$
 \item After performing the two previous steps for $x$, $y$ and $z$ redundant sequences are removed.
\end{enumerate}

\par
As an example the set of cells for the simplex $\mathcal{S} = \{ 3, 4, 5, 8\}$ in $\mathbf{C}$ are

\par
             $\{\{\{(5 \ 8)(4)(3)\}_x  \ ,            \ \{(8)(4)(3 \ 5)\}_y \ , \ \{(8)(4 \ 5)(3)\}_z\}, \\
\hspace*{1.7mm} \{\{(5 \ 8)(4)(3)\}_x  \ ,            \ \{(8)(4)(3 \ 5)\}_y \ , \ \{(8)(5)(3 \ 4)\}_z\}, \\
\hspace*{1.7mm} \{\{(5)(8)(3 \ 4)\}_x  \ ,            \ \{(8)(4)(3 \ 5)\}_y \ , \ \{(8)(4 \ 5)(3)\}_z\}, \\
\hspace*{1.7mm} \{\{(5 \ 8)(3 \ 4)\}_x \hspace{3mm} , \ \{(8)(4 \ 5)(3)\}_y \ , \ \{(8)(3 \ 4 \ 5)\}_z\}\}$ \hspace{7.2mm} (10)

\par
applying the above tranformation gives

\par
             $\{\{\{(5 \ 8)(3 \ 4)\}_x \ , \ \{(8)(4)(3 \ 5)\}_y \ , \ \{(8)(3 \ 4 \ 5)\}_z\}, \\
\hspace*{1.7mm} \{\{(5 \ 8)(3 \ 4)\}_x \ , \ \{(8)(4 \ 5)(3)\}_y \ , \ \{(8)(3 \ 4 \ 5)\}_z\}\}$ \hspace{9mm} (11)

(11) is a cone that contains (10), in $GDPS$ notation

\par
$\{\{\ose{1} 5 \
             8 \osd{1}
     \ose{2} 3 \
             4 \osd{2} \}_{x} \ , \
   \{\ose{1} 8 \osd{1} \
     \ose{2} 4 \
     \ose{3} 5 \osd{2} \
             3 \osd{3} \}_{y} \ , \
   \{\ose{1} 8 \osd{1} \
     \ose{2} 3 \
             4 \
             5 \osd{2} \}_{z}\}$ \hspace{8.6mm} (12)

\par
This operation is performed in order to determine the minimal set of ordinary cones that contain the cells from a given simplex.

\par
One aim of this note is to show that the set of rules described above give meaninful results, a detailed study of the decomposition of (9) will be the subject of a further communication. Here, we show the above procedure at work on a simpler case, that of the  first 13 $\alpha$-carbons from our example molecule, they have the $GDPS$ sequence

\par
$\{\{\ose{1}   5 \
     \ose{2}   8 \
     \ose{3}   6 \
     \ose{4}   9 \osd{1}
              11 \osd{2}
               7 \osd{3}
     \ose{5}   3 \
               4 \
              10 \osd{4}
              15 \
     \ose{6}  12 \osd{5} \
     \ose{7}  14 \osd{6}
              13 \osd{7}  \}_{x} \ , \\
\hspace*{1.7mm}
   \{\ose{1}  15 \osd{1} \
     \ose{2}  14 \osd{2} \
     \ose{3}  13 \osd{3} \
     \ose{4}  11 \
              12 \osd{4} \
     \ose{5}  10 \osd{5} \
     \ose{6}   9 \osd{6} \
     \ose{7}   8 \osd{7} \
     \ose{8}   7 \
     \ose{9}   4 \osd{8} \
     \ose{10}  5 \
               6 \osd{9}
               3 \osd{10} \}_{y} \ , \\
\hspace*{1.7mm}
   \{\ose{1}   7 \
               8 \
              10 \
     \ose{2}   6 \
               9 \
              11 \
              13 \
     \ose{3}  12 \osd{1} \
     \ose{4}  15 \osd{2}
              14 \osd{3}
               3 \
               4 \
               5 \osd{4} \}_{z}\}$ \hspace{49.3mm} (13)

\par
We can see a total of 7 sequences of permuting numbers in $x$, 10 in $y$ and 4 in $z$, our interest here is to see the realizable combinations of them obtained by merging compatible transformed sequences from different simplexes.

\par
A first example is from the $x_4$ sequence that has 6 numbers in it $(3\ 4\ 7\ 9\ 10\ 11)$, this means that the basic information about it, that is: how it connects with sequences in other dimensions, is contained in 15 simplexes, whose transformed $DPS$s are

\par
$\{\{(3\hspace*{1.7mm}4\hspace*{3.4mm}7\hspace*{3.4mm}9)\}_x \ ,
                        \ \{\hspace*{1.7mm}(9)(4\hspace*{3.4mm}7)\hspace*{1.7mm}(3)\}_y \ ,
                                          \ \{ (7\hspace*{3.4mm}9)(3\hspace*{3.4mm}4)\}_z\} \\
 \{\{(3\hspace*{1.7mm}4\hspace*{3.4mm}7\hspace*{1.7mm}10)\}_x \ ,
                        \ \{(10)(4\hspace*{3.4mm}7)\hspace*{1.7mm}(3)\}_y \ ,
                                          \ \{ (7\hspace*{1.7mm}10)(3\hspace*{3.4mm}4)\}_z\} \\
 \{\{(3\hspace*{1.7mm}4\hspace*{3.4mm}7\hspace*{1.7mm}11)\}_x \ ,
                        \ \{(11)(4\hspace*{3.4mm}7)\hspace*{1.7mm}(3)\}_y \ ,
                                          \ \{ (7\hspace*{1.7mm}11)(3\hspace*{3.4mm}4)\}_z\} \\
 \{\{(3\hspace*{1.7mm}4\hspace*{3.4mm}9\hspace*{1.7mm}10)\}_x \ ,
                        \ \{(10)(9)(4)\hspace*{1.7mm}(3)\}_y \ ,
                                          \ \{ (9\hspace*{1.7mm}10)(3\hspace*{3.4mm}4)\}_z\} \\
 \{\{(3\hspace*{1.7mm}4\hspace*{3.4mm}9\hspace*{1.7mm}11)\}_x \ ,
                        \ \{(11) (9)(4)\hspace*{1.7mm}(3)\}_y \ ,
                                          \ \{ (9\hspace*{1.7mm}11) (3\hspace*{3.4mm}4)\}_z\} \\
 \{\{(3\hspace*{1.7mm}4\hspace*{1.7mm}10\hspace*{1.7mm}11)\}_x \ ,
                        \ \{(11)(10)(4)(3)\}_y \ ,
                                          \ \{(10 \ 11) (3 \ 4)\}_z\} \\
 \{\{(3\hspace*{1.7mm}7\hspace*{3.4mm}9\hspace*{1.7mm}10)\}_x \ ,
                        \ \{(10)(9)(7)\hspace*{1.7mm}(3)\}_y \ ,
                                          \ \{ (7\hspace*{3.4mm}9\hspace*{1.7mm}10)(3)\}_z\} \\
 \{\{(3\hspace*{1.7mm}7\hspace*{3.4mm}9\hspace*{1.7mm}11)\}_x \ ,
                        \ \{(11)(9)(7)\hspace*{1.7mm}(3)\}_y \ ,
                                          \ \{ (7\hspace*{3.4mm}9\hspace*{1.7mm}11)(3)\}_z\} \\
 \{\{(3\hspace*{1.7mm}7\hspace*{1.7mm}10\hspace*{1.7mm}11)\}_x \ ,
                         \ \{(11)(10)(7)(3)\}_y \ ,
                                          \ \{ (7 \ 10 \ 11)(3)\}_z\} \\
 \{\{(3\hspace*{1.7mm}9\hspace*{1.7mm}10\hspace*{1.7mm}11)\}_x \ ,
                         \ \{(11)(10)(9)(3)\}_y \ ,
                                          \ \{ (9 \ 10 \ 11)(3)\}_z\} \\
 \{\{(4\hspace*{1.7mm}7\hspace*{3.4mm}9\hspace*{1.7mm}10)\}_x \ ,
                         \ \{(10)\hspace*{1.7mm}(9) (4\hspace*{3.4mm}7)\}_y \ ,
                                          \ \{ (7\hspace*{3.4mm}9\hspace*{1.7mm}10)(4)\}_z\} \\
 \{\{(4\hspace*{1.7mm}7\hspace*{3.4mm}9\hspace*{1.7mm}11)\}_x \ ,
                         \ \{(11)\hspace*{1.7mm}(9) (4\hspace*{3.4mm}7)\}_y \ ,
                                          \ \{ (7\hspace*{3.4mm}9\hspace*{1.7mm}11)(4)\}_z\} \\
 \{\{(4\hspace*{1.7mm}7\hspace*{1.7mm}10\hspace*{1.7mm}11)\}_x \ ,
                         \ \{(11)(10)(4\hspace*{3.4mm}7)\}_y \ ,
                                          \ \{ (7 \ 10 \ 11)(4)\}_z\} \\
 \{\{(4\hspace*{1.7mm}9\hspace*{1.7mm}10\hspace*{1.7mm}11)\}_x \ ,
                         \ \{(11)(10)(9)(4)\}_y \ ,
                                          \ \{ (9 \ 10 \ 11)(4)\}_z\} \\
 \{\{(7\hspace*{1.7mm}9\hspace*{1.7mm}10\hspace*{1.7mm}11)\}_x \ ,
                         \ \{(11)(10)(9)(7)\}_y \ ,
                                          \ \{ (7\hspace*{3.4mm}9 \ 10 \ 11)\}_z\}$ \hspace{27mm} (14)

\par
obviously the sequences from these 15 cones are all compatible with those from adjacent simplexes, so they can all be merged to give the result

$\{\{(3\ 4\ 7\ 9\ 10\ 11)\}_x\ ,\ \{(11)(10)(9)(4\ 7)(3)\}_y\ ,\ \{(7\ 9\ 10\ 11)(3\ 4)\}_z\}$ \hspace{8mm} (15)

\par
in which $x_4$ appears to be connected with sequences $z_1$ and $z_4$, or at least with fragments of them.
Contrasting with (15) the sequence $z_1$, of length 8, after merging up the $DPS$s from 70 simplexes we end with a total of 7 cones

$\{\{(8\ 9)(6\ 7\ 11)(10)(12)(13)\}_x\   ,\ \{(13)(11\ 12)(7)(10)(9)(8)(7)(6)\}_y\ ,\ \{(6\ 7\ 8\ 9\ 10\ 11\ 12\ 13)\}_z\}\\
 \{\{(8\ 9\ 11)(6\ 7)(10)(12)(13)\}_x\   ,\ \{(13)(11\ 12)(7)(10)(9)(8)(7)(6)\}_y\ ,\ \{(6\ 7\ 8\ 9\ 10\ 11\ 12\ 13)\}_z\}\\
 \{\{(8)(6\ 7\ 9\ 11)(10)(12)(13)\}_x\   ,\ \{(13)(11\ 12)(7)(10)(9)(8)(7)(6)\}_y\ ,\ \{(6\ 7\ 8\ 9\ 10\ 11\ 12\ 13)\}_z\}\\
 \{\{(8\ 11)(6\ 7\ 9)(10)(12)(13)\}_x\   ,\ \{(13)(11\ 12)(7)(10)(9)(8)(7)(6)\}_y\ ,\ \{(6\ 7\ 8\ 9\ 10\ 11\ 12\ 13)\}_z\}\\
 \{\{(6\ 8\ 11)(7\ 9\ 10)\ (12)(13)\}_x\ ,\ \{(13)(11\ 12)(7)(10)(9)(8)(7)(6)\}_y\ ,\ \{(6\ 7\ 8\ 9\ 10\ 11\ 12\ 13)\}_z\}\\
 \{\{(6\ 8)(7\ 9\ 10\ 11)\ (12)(13)\}_x\ ,\ \{(13)(11\ 12)(7)(10)(9)(8)(7)(6)\}_y\ ,\ \{(6\ 7\ 8\ 9\ 10\ 11\ 12\ 13)\}_z\}\\
 \{\{(6\ 8\ 9\ 11)(7\ 10)\ (12)(13)\}_x\ ,\ \{(13)(11\ 12)(7)(10)(9)(8)(7)(6)\}_y\ ,\ \{(6\ 7\ 8\ 9\ 10\ 11\ 12\ 13)\}_z\}$

$z_1$ appears to be connected with $x_2$, $x_3$ and $x_4$, but there is also 4 cones that combine with fragments from the main sequences.
This reveals details of how $3D$ structures are organized inside the main cone: only some of the $DPS$s in one dimension connect with those from another dimension. The exception in this example are the segments from the $y$ dimension: there is so few structure in them that they appear to combine with any other segment from $x$ and $y$.

\par
The remaining segments $x_1$, $x_5$ and $z_4$ give

$\{\{(5\ 6\ 8\ 9)\}_x\ ,\ \{(9)(8)(5\ 6)\}_y\ ,\ \{(6\ 8\ 9)(5)\}_z\}$  \hspace{38mm} ($x_1$)

$\{\{(3\ 4\ 10\ 12\ 15)\}_x\ ,\ \{(15)(12)(10)(4)(3)\}_y\ ,\ \{(10\ 12\ 15)(3\ 4)\}_z\}$  \hspace{8mm} ($x_5$)

$\{\{(5)(3\ 4\ 15)(14)\}_x\ ,\ \{(15)(14)(4)(3\ 5)\}_y\ ,\ \{(3\ 4\ 5\ 14\ 15)\}_z\}$ \newline
$\{\{(5)(3\ 4\ 15)(14)\}_x\ ,\ \{(15)(14)(4\ 5)(3)\}_y\ ,\ \{(3\ 4\ 5\ 14\ 15)\}_z\}$ \hspace{14mm} ($z_4$)

\vspace*{4mm}
\begin{center}
{\scshape VI. Conclusion}
\end{center}

\par
The $GDPS$ (7) is a global approximation that sets the bounds of the region from $CS$ in which the system evolves. These bounds are set independently for the $x$, $y$ and $z$ coordinates of the molecule, and consequently they are not correlated: this means that much of the volume enclosed by (7) does not correspond to realizable $3D$-structures.

\par
On the other hand the graph of cells $\mathbf{G}$ allows to exactly enumerate the set of visited cells in conformational space, but this possibility is probably algorithmically hopeless.

\par
The approach presented in this paper consists in using the information contained in $\mathbf{G}$ to derive bounds from (7) that are correlated in $x$, $y$ and $z$, and to progressively narrow these bounds around interesting regions.

\par
The importance of a structure like (7), and the $DPS$s that are derived from them, is not only the precision they can attain in delimiting conformational space,
but the fact that they possess a \textbf{graphical structure} and connected graphs always have a metric: we can measure the distance between two nodes as the length of the minimal path that joins them.
This opens the very real possibility of measuring and enumerating distances between points and conformations, so that a combinatorial hamiltonian can be built on these structures. And this should be the next phase of the present work.

\newpage
\vspace*{4mm}
\begin{center}
{\scshape References}
\end{center}
\begin{itemize}

\item[1.]  J. Gabarro-Arpa,
          "A central partition of molecular conformational space. I. Basic structures",
          {\it Comp. Biol. and Chem.}
          {\bf 27}, 153-159 (2003).

\item[2.]  J. Gabarro-Arpa,
          "A central partition of molecular conformational space. II. Embedding 3D-structures",
          {\it Proceedings of the 26th Annual International Conference of the IEEE EMBS},
          San Francisco, 3007-3010 (2004).

\item[3.]  J. Gabarro-Arpa,
          "Combinatorial determination of the volume spanned by a molecular system in conformational space",
          {\it Lecture Series on Computer and Computational Sciences}
          {\bf 4}, 1778-1781 (2005).

\item[4.]  J. Gabarro-Arpa,
          "A central partition of molecular conformational space. III.
           Combinatorial determination of the volume spanned by a molecular system in conformational space",
          {\it Journal of Mathematical Chemistry }
           DOI 10.1007 s10910-006-9079-8.url (2006).

\item[5.]  J. Gabarro-Arpa,
          "A Central Partition of Molecular Conformational Space. IV. Extracting information from the graph of cells",
           arXiv: physics/061108v2, submitted for publication (2007).

\item[6.] S. Fomin and N. Reading, "Root systems and generalized associahedra",
          math.CO/0505518 (2005).

\item[7.]  A. Bjorner, M. las Vergnas, B. Sturmfels, N. White,
          "Oriented Matroids",
           Cambridge, UK, Cambridge University Press, 
           sect. {\bf 2} (1993).

\item[8.]  M. Marquart, J. Walter, J. Deisenhofer, W. Bode, R. Huber,
            "The geometry of the reactive site and of the peptide groups in trypsin,
             trypsinogen and its complexes with inhibitors",
            {\it Acta Crystallogr. Sect. B},
            {\bf 39}, 480-490 (1983).

\item[9.]  J. Gabarro-Arpa, R. Revilla,
          "Clustering of a molecular dynamics trajectory with a Hamming distance",
          \textit{Comp. and Chem. },
          \textbf{24}, 693-698 (2000).

\end{itemize}

\end{document}